\documentclass[aps,a4paper,superscriptaddress,nofootinbib,onecolumn]{revtex4-1}
\usepackage{amsmath,amssymb,gensymb,graphicx,multirow,dcolumn,bm,latexsym,soul,nicefrac,acronym}
\usepackage{appendix}
\usepackage[colorlinks,linkcolor=blue,citecolor=blue,urlcolor=blue ]{hyperref}
\usepackage{color}
\usepackage{cancel}
\usepackage{acronym}
\usepackage{subfigure}
\usepackage{appendix}
\usepackage{longtable}
\usepackage{footnote}
\usepackage[normalem]{ulem}
\newcommand{\mnras}{Monthly Notices of the Royal Astronomical Society}

\newcommand{\be}{\begin{equation}}
\newcommand{\ee}{\end{equation}}
\newcommand{\bea}{\begin{eqnarray}}
\newcommand{\eea}{\end{eqnarray}}
\newcommand{\bw}{\begin{widetext}}
\newcommand{\ew}{\end{widetext}}
\newcommand{\nn}{\nonumber}

\newcommand{\eq}[1]{Eq.~(\ref{#1})}
\newcommand{\fig}[1]{Fig.~\ref{#1}}
\newcommand{\tab}[1]{Table.~\ref{#1}}

\newcommand{\HNU}{School of Physics, Henan Normal University, Xinxiang 453007, China}

\newcommand{\TRC}{MOE Key Laboratory of TianQin Mission, TianQin Research Center for Gravitational Physics $\&$  School of Physics and Astronomy, Frontiers Science Center for TianQin, CNSA Research Center for Gravitational Waves, Sun Yat-sen University (Zhuhai Campus), Zhuhai 519082, China}

\begin{document}
\title{Probing Dark Matter's Gravitational Effects Locally with TianQin}
\author{Zheng-Cheng Liang}
%\email{liangzhch7@mail.sysu.edu.cn}
\affiliation{\HNU}
\author{Fa-Peng Huang}
\email{Corresponding author: huangfp8@sysu.edu.cn}
\affiliation{\TRC}
\author{Xuefeng Zhang}
%\email{Corresponding author: zhangxf38@mail.sysu.edu.cn}
\affiliation{\TRC}
\author{Yi-Ming Hu}
\email{Corresponding author: huyiming@sysu.edu.cn}
\affiliation{\TRC}

\date{\today}

\begin{abstract}
In this study, we explore the potential of using TianQin missions to probe the local gravitational effects of dark matter. 
The TianQin project plans to launch satellites at both low and high orbits. 
High-precision orbit determination is expected to aid in detecting Earth's gravity or gravitational waves. 
By comparing the derived masses in low and high orbits, it is possible to constrain the amount of dark matter between the two spheres, hence placing a local constraint on dark matter's gravitational effect. 
Our results show the capability of TianQin in detecting the density of dark matter around Earth, with an ultimate sensitivity to a value of $10^{-8}\,\,{\rm kg\,\,m^{-3}}$. 
This detection limit surpasses the estimated bounds for the solar system and the observation results for our Galaxy by approximately 7 and 14 orders of magnitude, respectively. 
\end{abstract}

\keywords{}

\pacs{04.25.dg, 04.40.Nr, 04.70.-s, 04.70.Bw}

%%%%%%%%%%%%%%%%%%%%%%%%%%%%%%%%%%%%%%%%%%%%%%%%%%%%%%%%%%%%%%%%
\maketitle
\acrodef{GW}{gravitational-wave}
\acrodef{LDM}{local dark mater}
\acrodef{CMB}{cosmic microwave background}
\acrodef{LIGO}{Laser interferometry Gravitational-Wave Observatory}
\acrodef{PTA}{Pulsar Timing Arrays}
\acrodef{TQ}{TianQin}
\acrodef{POD}{precision orbit determination}
\acrodef{LISA}{Laser Interferometer Space Antenna}
\acrodef{LEO}{low Earth orbit}
\acrodef{LLR}{lunar laser ranging}

%%%%%%%%%%%%%%%%%%%%%%%%%%%%%%%%%%%%%%%%%%%%%%%%%%%%%%%%%%%%%%%%
%%%% ±êÌâÒ³œáÊø %%%%%%%%%%%%%%%%%%%%%%%%%%%%%%%%%%%%%%%%%%%%%%%%
%%%%%%%%%%%%%%%%%%%%%%%%%%%%%%%%%%%%%%%%%%%%%%%%%%%%%%%%%%%%%%%%
%%%% µÚÒ»œÚ¿ªÊŒ %%%%%%%%%%%%%%%%%%%%%%%%%%%%%%%%%%%%%%%%%%%%%%%%
%%%%%%%%%%%%%%%%%%%%%%%%%%%%%%%%%%%%%%%%%%%%%%%%%%%%%%%%%%%%%%%%
\section{Introduction}
Dark matter, a concept of immense significance in cosmology and astronomy, is a fascinating topic. 
This hypothetical substance plays a critical role in our understanding of the universe~\cite{Bertone:2016nfn}. 
Furthermore, it is widely recognized that dark matter cannot be composed of ordinary particles found within the framework of the standard model~\cite{Bertone:2004pz}. 
Consequently, the pursuit of dark matter detection holds paramount importance across various scientific disciplines~\cite{Mayet:2016zxu}. 

The existence of dark matter has been inferred from its gravitational effects observed in astronomical observations~\cite{DAmico:2009tep}. 
Measurements of the \ac{CMB} have provided valuable insights into the density of dark matter in the Universe, yielding a value of $2\times10^{-25}{\rm\,\,kg\,\,m^{-3}}$~\cite{Planck:2018vyg}. 
Further examinations of the vertical kinematics of disc stars and the rotation curve of the Galaxy have led to an estimation of the current density of \ac{LDM} in the Galaxy, denoted as $\rho_{\rm DM}=5\times10^{-22}{\rm \,\,kg\,\,m^{-3}}$~\cite{deSalas:2020hbh}. 
Regarding our solar system, utilizing the EPM2011 planetary ephemeris and position observations from planets and spacecrafts, an upper limit for the density of \ac{LDM} has been established at $1\times10^{-14}{\rm \,\,kg\,\,m^{-3}}$~\cite{2013AstL...39..141P,2013MNRAS.432.3431P}. 
Additional constraints on this value have been obtained through extensive tracking and modeling of the trajectory of the asteroid (101955) Bennu from the OSIRIS-REx mission, yielding a refined estimate of $3\times10^{-15}{\rm \,\,kg\,\,m^{-3}}$~\cite{Tsai:2022jnv}. 
In addition, \ac{GW} detectors including \ac{LISA}, Taiji, and \ac{PTA} promise to provide another approach to constrain the density of \ac{LDM}~\cite{Cerdonio:2008un,Porayko:2018sfa,PPTA:2022eul,NANOGrav:2023hvm,EPTA:2023xxk,EuropeanPulsarTimingArray:2023egv,Yao:2024fie,Yu:2024enm}. 

While significant findings have been made, it is crucial to acknowledge that particle physics experiments conducted near the Earth have not yet yielded direct evidence of dark matter particles~\cite{Liu:2017drf,Hochberg:2022apz}. 
However, the presence of dark matter particles in the vicinity of the Earth can lead to variations in the gravitational effects at different radii within Earth's dark matter halo. 
These variations can serve as potential indications of the existence of \ac{LDM}. 
Consequently, alongside particle experiments, exploring the gravitational effects becomes essential in the quest to detect \ac{LDM} around the Earth. 

In this paper, we investigate the gravitational effects of \ac{LDM} near the Earth by leveraging the \ac{TQ} project. 
The \ac{TQ} project is primarily dedicated to the space-borne \ac{GW} detection~\cite{TianQin:2015yph}. 
Its implementation involves launching satellites at two different orbital altitudes, roughly $200\,\,\rm km$ and $10^{5}\,\,\rm km$~\cite{TianQin:2020hid}. 
This strategic arrangement offers a valuable opportunity to estimate the masses of gravitational sources by utilizing the orbital radius and period measurements of the satellites at different altitudes. 
By comparing the derived masses, we can effectively detect the presence of \ac{LDM}. 

This paper is organized as follows. 
In Sec.~\ref{sec:Methodology}, we present our methodology for detecting the \ac{LDM} near the Earth. 
Then, we apply our methodology to the TianQin project in Sec.~\ref{sec:TQcase}. 
We provide a summary of our study in Sec.~\ref{sec:summary}.

\section{Methodology}\label{sec:Methodology}
To start, this section involves the methodology for detecting \ac{LDM} near the Earth through its gravitational effects. 
In our approach, we assume that the \ac{LDM} surrounding the Earth forms a homogeneous diffuse background of particles. 
With this assumption, we can derive an expression that describes the gravitational effects experienced by a satellite orbiting at the radius $R$ and period $T$:
\be
\label{eq:GM}
G(M_{\rm E}+M_{\rm DM})=\frac{4\pi^{2}R^{3}}{T^{2}}.
\ee
The derived expression incorporates fundamental parameters such as the gravitational constant $G$, the mass of the Earth $M_{\rm E}$, and the mass of \ac{LDM} within a sphere of radius $R$, which we denote as $M_{\rm DM}$. 
It is essential to note that our analysis focuses on circular orbits and neglects the mass of the satellite itself. 

\begin{figure}[htbp]
	\centering
	\includegraphics[height=8cm]{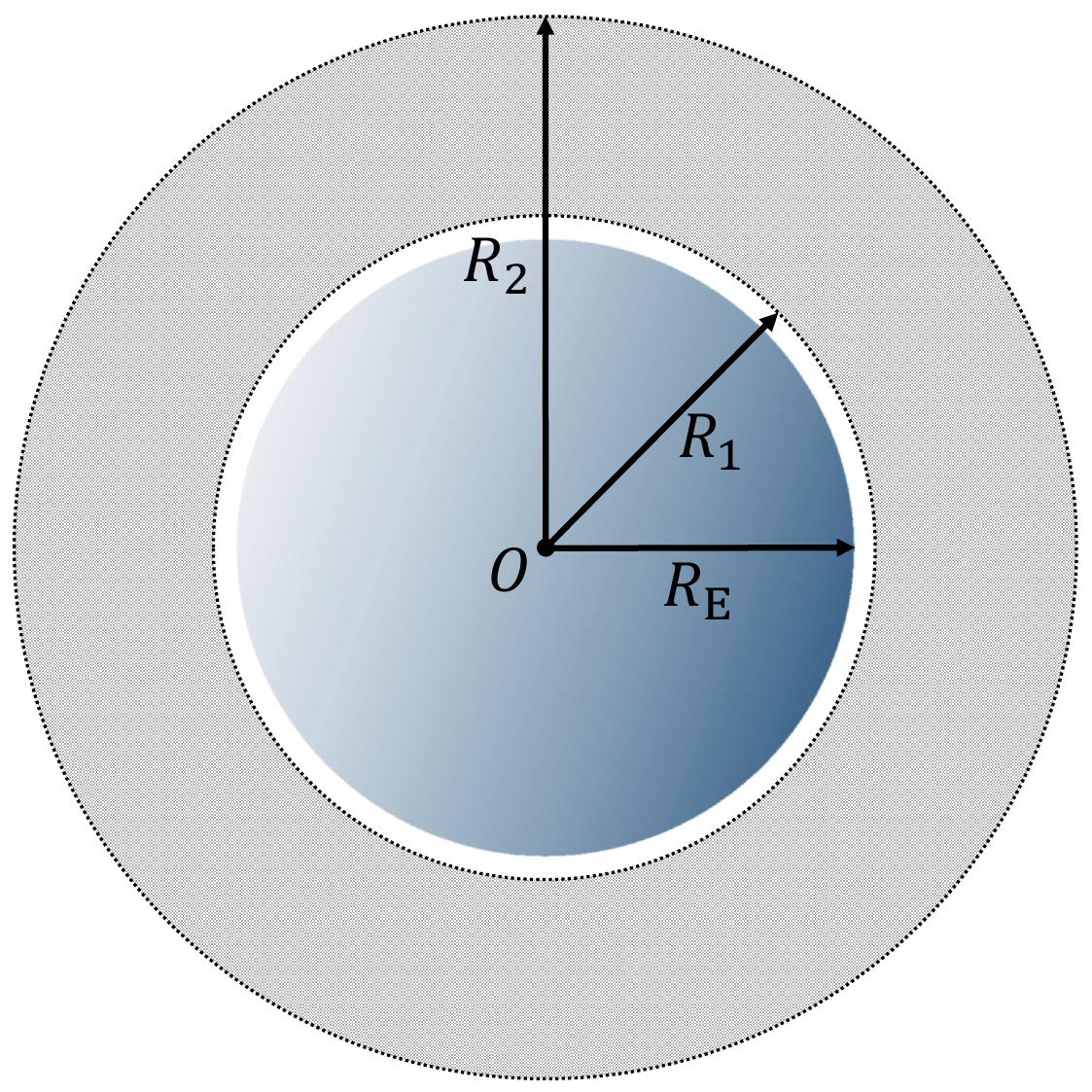}
	\caption{Schematic diagram of \ac{LDM} detection. The radius of the Earth $R_{\rm E}\simeq6400\,\,\rm km$, the shadowed part is represented as the measured dark matter region.}
	\label{fig:LDM_detection}
\end{figure}

In \fig{fig:LDM_detection}, we can observe a viable approach to address the potential impact of measurement errors in the Earth mass.  This approach involves using two satellites with different orbital radii, denoted as $R_{1}$ and $R_{2}$. 
Furthermore, the corresponding orbital periods of these satellites are represented by $T_{1}$ and $T_{2}$. 
By employing this setup, it becomes possible to calculate the density of \ac{LDM} within the shell situated between these two satellites:
\be
\label{eq:rho}
\rho_{\rm DM}=
\frac{3\pi(R_{2}^{3}T_{1}^{2}-R_{1}^{3}T_{2}^{2})}
{G(R_{2}^{3}-R_{1}^{3})T_{1}^{2}T_{2}^{2}}.
\ee
For comprehensive analytical details, readers may refer to the complete derivation process documented in Appendix~\ref{appen:der}.
To account for the errors in orbital radius and period measurements, it is crucial to incorporate the error propagation formula when calculating the measurement error of the \ac{LDM} density. 
Assuming all parameters are independent, one can determine the measurement error for the \ac{LDM} density, which corresponds to the upper limit of detection, using the following formula:
\bea
\label{eq:sigma}
\sigma_{\rho_{\rm DM}}=
\sqrt{\sum_{i}\left(\frac{\partial \rho_{\rm DM}}{\partial X_{i}}\right)^2\sigma^{2}_{X_{i}}},
\eea
where the parameter group $X=\{R_{1},R_{2},T_{1},T_{2}\}$. 
Substituting \eq{eq:rho} to \eq{eq:sigma}, we have
\bea
\label{eq:Drho}
\nn
\frac{\partial \rho_{\rm DM}}{\partial R_{1}}&=&
\frac{A_{\rm R}}{R_{1}},\\
\nn
\frac{\partial \rho_{\rm DM}}{\partial R_{2}}&=&
-\frac{A_{\rm R}}{R_{2}},\\
\nn
\frac{\partial \rho_{\rm DM}}{\partial T_{1}}&=&
\frac{A_{\rm T_{1}}}{T_{1}},\\
\frac{\partial \rho_{\rm DM}}{\partial T_{2}}&=&
-\frac{A_{\rm T_{2}}}{T_{2}},
\eea
where 
\bea
\label{eq:A_R}
\nn
A_{\rm R}&=&\frac{9\pi R_{1}^3 R_{2}^3 (T_{2}^2-T_{1}^2)}{G (R_{2}^3-R_{1}^3)^2T_{1}^2 T_{2}^2},    \\
\nn
A_{\rm T_{1}}&=&\frac{6 \pi R_{1}^3}{G (R_{2}^3-R_{1}^3) T_{1}^2},\\
A_{\rm T_{2}}&=&\frac{6 \pi R_{2}^3}{G(R_{2}^3-R_{1}^3)T_{2}^2}.
\eea
As derived from~\eq{eq:GM}, when accounting for the fact that the mass of dark matter is negligible compared to the Earth' mass, we observe that the difference between $R_{1}^{3}/T_{1}^{2}$ and $R_{2}^{3}/T_{2}^{2}$ is significantly smaller than either term individually. 
This relationship can be mathematically represented as:  $R_{2}^{3}/T_{2}^{2}=R_{1}^{3}/T_{1}^{2}+o(R_{1}^{3}/T_{1}^{2})$, where $o(...)$ term captures higher-order corrections. 
Given the above conditions, \eq{eq:A_R} implies
\bea
\label{eq:A_R1}
\nn
A_{\rm R}&\propto& {\rm const},\\
A_{\rm T_{1}},A_{\rm T_{2}}&\propto& \frac{1}{R_{2}^{3}-R_{1}^{3}}.
\eea
Referring back to~Eqs.~(\ref{eq:sigma}) and~(\ref{eq:Drho}), we observe that measurement uncertainty from the orbital radius $R_{i}$ contributes an error term proportional to $\sigma_{R_{i}}/R_{i}$. 
Similarly, the uncertainty from period $T_{i}$ scales as $\sigma_{T_{i}}/T_{i}$. 
Nevertheless, the impact of period uncertainty can be mitigated by increasing the radial separation $R_{2}-R_{1}$ between the two measurement orbits.

\section{TianQin case}\label{sec:TQcase}
The TQ project aims to detect \acp{GW} within the $\rm mHz$ frequency range~\cite{TianQin:2015yph}. 
It involves the deployment of three Earth-orbiting satellites with an orbital radius of approximately $10^{5}\,\,\rm km$. 
These satellites form an equilateral triangle constellation positioned perpendicular to the ecliptic plane. 
To ensure the successful implementation of the \ac{TQ} project and the maturity of critical technologies, a roadmap named the 0123 plan was adopted in 2015. 
This roadmap provides a strategic framework consisting of several steps that need to be followed~\cite{Luo:2020bls,TianQin:2020hid}: step 0 involves acquiring the capability to obtain high-precision orbit information for satellites in the \ac{TQ} orbit through \ac{LLR} experiments; 
step 1 focuses on the single satellite mission aimed at testing and demonstrating the maturity of inertial reference technology; 
step 2 entails a mission with a pair of satellites to evaluate and showcase the maturity of inter-satellite laser interferometry technology; 
the final step, step 3, marks the launch of all three satellites to form the space-borne \ac{GW} detector known as \ac{TQ}. 
As part of the TQ project, step 2 involves the deployment of the TianQin-2 satellite in a \ac{LEO} at an orbital altitude of $200\,\,\rm km$. 
In Step 3, the TianQin-3 satellite will be positioned at an orbital radius of $10^{5}\,\,\rm km$. 
Their respective orbital periods are approximately 1.5 hours and 3.64 days.

In this work, we aim to explore the upper measurement limit for LDM density with the TQ project, as constrained by the precision of orbital radius and period measurements. 
Multiple \ac{POD} methods exist for measuring orbital radius, including global navigation satellite system (GNSS)~\cite{2018ChJAn}, deep space network (DSN)~\cite{2019AdSpR}, and satellite laser ranging (SLR)~\cite{2001SGeo}. 
Among these methods, SLR offers a higher accuracy ranging system, determining the satellite-to-station distance by precisely measuring the flight time of laser pulses. 
For the TianQin project, POD measurement errors are projected to be on the order of $10^{-3}\,\,\rm m$ for TianQin-2 and $10^{-1}\,\,\rm m$ for TianQin-3, respectively~\cite{An:2022xsi}. 
Furthermore, the orbital period measurement can utilize relative clock alignment, achieving an estimated error level of $2\times10^{-11}\,\,\rm s$~\cite{2020AdSpR..66..469B,An:2022xsi}.

\begin{table}[ht]
	\begin{center}
		\caption{Measurement error contribution of different parameters.}
		\label{tab:sigma}
		\setlength{\tabcolsep}{1.8mm}
		\renewcommand\arraystretch{1.3}
		\begin{tabular}{*{5}{|c}|}
			\hline
			$X_{i}$  & $R_{1}$ & $R_{2}$ & $T_{1}$ & $T_{2}$  \\
			\hline
			$|\frac{\partial \rho_{\rm DM}}{\partial X_{i}}\sigma_{X_{i}}|$  & \multirow{2}*{$1.5\times10^{-9}$} & \multirow{2}*{$9.7\times10^{-9}$} & \multirow{2}*{$3.7\times10^{-14}$} & \multirow{2}*{$1.8\times10^{-16}$}  \\
			$(\rm kg\,\,m^{-3})$  & & & & \\
			\hline
		\end{tabular}
	\end{center}
\end{table}

\begin{figure}[h]
	\centering
	\includegraphics[height=8cm]{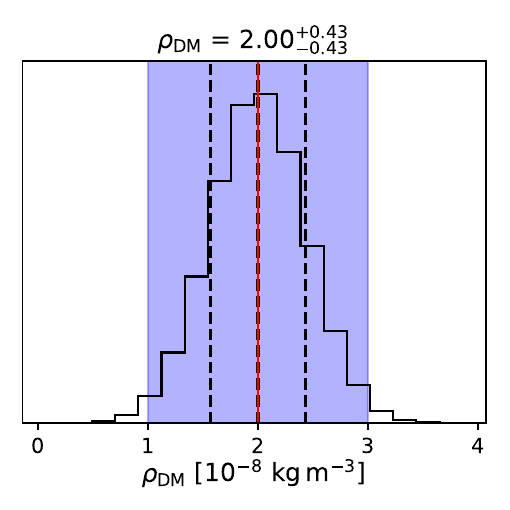}
	\caption{Monte Carlo simulations for the LDM density. Vertical dashed lines on the distribution mark the quantiles [16\%, 50\%, 84\%], with the red line representing the true value. The blue band covers the theoretical measurement uncertainty.}
	\label{fig:rho_DM}
\end{figure}

By substituting the aforementioned measurement error values into~\eq{eq:sigma}, we can calculate the upper limit of the \ac{LDM} density, which is found to be $\sigma_{\rho_{\rm DM}}=1\times10^{-8}\,\,{\rm kg\,\,m^{-3}}$. 
\tab{tab:sigma} provides a numerical breakdown of each parameter's contribution to the estimation of $\sigma_{\rho_{\rm DM}}$, further emphasizing the dominant role played by certain parameters' measurement accuracy in determining the density error. 
In this case, the orbital radius emerges as a crucial factor that requires improved precision, as the orbital radius measurement has an accuracy that is approximately 5 orders of magnitude less precise compared to the orbital period measurement. 
This disparity suggests that significant improvements in current \ac{POD} technology would substantially contribute to the \ac{LDM} detection.

For numerical validation, we conducted Monte Carlo simulations with the LDM density $\rho_{\rm DM}$ fixed at $2\times10^{-8}\,\,{\rm kg\,\,m^{-3}}$. 
Using the true values of parameter group $\{R_{1},R_{2},T_{1},T_{2}\}$ as distribution means and the measurement uncertainties listed in \tab{tab:sigma} as standard deviations, we generated one million Gaussian-distributed random samples for each parameter. 
As shown in~\fig{fig:rho_DM}, the resulting distribution of the $\rho_{\rm DM}$ converges to a Gaussian distribution characterized by $2.00\pm0.43 \times 10^{-8}\,\,{\rm kg\,\,m^{-3}}$, where the 1-$\sigma$ confidence interval is entirely contained within the theoretical measurement uncertainty bound. 
This confirms the experimental feasibility of detecting the LDM density at the $10^{-8}\,\,{\rm kg\,\,m^{-3}}$ scale through our proposed measurement approach.

\section{Summary}\label{sec:summary}
%Detecting \ac{LDM} is integral to understanding its properties and presence, with far-reaching implications across various scientific domains. 
%It contributes to exploring the origins of the universe, investigating galaxy formation, and advancing fundamental physics research. 
This study examined the feasibility of utilizing Earth-orbiting satellites at different orbital altitudes to detect \ac{LDM} around the Earth. 
Our results suggested that deploying satellites in both \ac{LEO} and higher orbital altitudes can significantly enhance the detection capability to \ac{LDM}, with particular emphasis on increasing the orbital altitude of the latter satellite. 

For the TianQin-2 and TianQin-3 satellites employed in the \ac{TQ} project, we calculated the detection limit for \ac{LDM} density near the Earth, yielding a value of $1\times10^{-8}\,\,{\rm kg\,\,m^{-3}}$. 
When compared with the observation results of our Galaxy and the estimated constraints of the solar system, this value falls significantly short of the required detection threshold. 
Nevertheless, it is crucial to acknowledge that advancements in the \ac{POD} technology can potentially enhance this detection limit. 
Furthermore, our method---unlike \ac{GW} detection---relies uniquely on POD for \ac{LDM} detection in conjunction with TianQin. Due to this fundamental methodological difference, our approach is not directly applicable to LISA or Taiji.

While our current analysis assumes an isotropic dark matter distribution for simplicity, we acknowledge that the LDM halo may exhibit anisotropies similar to those at galactic scales~\cite{Bozorgnia:2013pua,Bozorgnia:2019mjk}. 
However, constraining such anisotropic features in the vicinity of Earth presents significant observational challenges. 
The region surrounding Earth spans scales orders of magnitude smaller than characteristic LDM halo structures in the Milky Way ($\rm kpc$) or even within the solar system (A.U.). 
This fundamental scale disparity limits our ability to resolve potential subdominant anisotropic contributions, which would require at least an order of magnitude better sensitivity to disentangle from the dominant isotropic background.

\appendix

\section{Corresponding derivation for the LDM density}\label{appen:der}
In terms of~\eq{eq:GM}, one can derive the enclosed mass profiles for two concentric spherical regions:
\bea
\nn
\label{eq:M_tot}
M_{\rm E}+M_{\rm DM_{1}}&=&\frac{4\pi^{2}R_{1}^{3}}{G T_{1}^{2}},\\
M_{\rm E}+M_{\rm DM_{2}}&=&\frac{4\pi^{2}R_{2}^{3}}{G T_{2}^{2}},
\eea
where $M_{{\rm DM}_{i}}$ represents the dark matter mass within radius $R_{i}$. 
By subtracting the first and second lines of~\eq{eq:M_tot}, the mass of the Earth cancels out, allowing direct determination of the LDM mass within the spherical shell region $R_1<r<R_2$:
\be
\label{eq:deltaM}
\Delta M_{\rm DM}=
\frac{4\pi^{2}}{G}\left(\frac{R_{2}^{3}}{T_{2}^{2}}-\frac{R_{1}^{3}}{T_{1}^{2}}\right).
\ee
The corresponding shell volume is given by standard spherical geometry:
\be
\label{eq:deltaV}
\Delta V_{\rm DM}=\frac{4}{3}\pi\left(R_{2}^{3}-R_{1}^{3}\right).
\ee
Combining Eqs.~(\ref{eq:deltaM}) and (\ref{eq:deltaV}) yields the LDM density $\rho_{\rm DM}$ as specified in~\eq{eq:rho}.

\begin{acknowledgments}
This work has been supported by the National Key Research and Development Program of China (No. 2023YFC2206704), the National Key Research and Development Program of China (No. 2020YFC2201400), and the Natural Science Foundation of China (Grant No. 12173104). 
Z.C.L. is supported by the Guangdong Basic and Applied Basic Research Foundation (Grant No. 2023A1515111184). 
X. Z. is supported by the Natural Science Foundation of China (Grant No. 12373116). 
\end{acknowledgments}

%%%%%%%%%%%%%%%%%%%%%%%%%%%%%%%%%%%%%%%%%%%%%%%%%%%%%%%%%%%%%%%%
%%%% ²Î¿¼ÎÄÏ×¿ªÊ¼ %%%%%%%%%%%%%%%%%%%%%%%%%%%%%%%%%%%%%%%%%%%%%%
%%%%%%%%%%%%%%%%%%%%%%%%%%%%%%%%%%%%%%%%%%%%%%%%%%%%%%%%%%%%%%%%
%\normalem
\bibliographystyle{apsrev4-1}
%%%%%%%%%%%%%%%%%%%%%%%%%%%%%%%%%%%%%%%%%%%%%%%%%%%%%%%%%%%%%%%%
%%% ½«²Î¿¼ÎÄÏ×Ìí¼Óµ½ reference.bib ÎÄ¼þÀï£¬ÔÚÕâÀïµ÷ÓÃ %%%%%%
%%%%%%%%%%%%%%%%%%%%%%%%%%%%%%%%%%%%%%%%%%%%%%%%%%%%%%%%%%%%%%%%
\bibliography{DM}

%%%%%%%%%%%%%%%%%%%%%%%%%%%%%%%%%%%%%%%%%%%%%%%%%%%%%%%%%%%%%%%%
%%%% ²Î¿¼ÎÄÏ×½áÊø %%%%%%%%%%%%%%%%%%%%%%%%%%%%%%%%%%%%%%%%%%%%%%
%%%%%%%%%%%%%%%%%%%%%%%%%%%%%%%%%%%%%%%%%%%%%%%%%%%%%%%%%%%%%%%%
\end{document}